\documentstyle[twdis98,psfig]{article}

\newcommand{\piz}{$\pi^{\circ}$}
\newcommand{\ptpi}{\mbox{$p_{t,\pi}$}}
\newcommand{\ptcp}{\mbox{$p_{t,cp}$}}
\newcommand{\pcp}{\mbox{$p_{cp}$}}
\newcommand{\pt}{\mbox{$p_{t}$}}
\newcommand{\etpi}{\mbox{$E_{t,\pi}$}}
\newcommand{\epi}{\mbox{$E_{\pi}$}}
\newcommand{\ecp}{\mbox{$E_{cp}$}}
\newcommand{\xpi}{\mbox{$x_{\pi}$}}
\newcommand{\xcp}{\mbox{$x_{cp}$}}
\newcommand{\xall}{\mbox{$x_{i}$}}
\newcommand{\epro}{\mbox{$E_{proton}$}}
\newcommand{\thpi}{\mbox{$\Theta_{\pi}$}}
\newcommand{\thcp}{\mbox{$\Theta_{cp}$}}
\newcommand{\Qsq}{\mbox{$Q^2$}}
\newcommand{\GeVsq}{\mbox{${\rm ~GeV}^2~$}}

\begin{document}

\title{Forward {\piz}-Meson and Charged Particle Production in Deep
  Inelastic Scattering at low Bjorken-$x$}

\author{Thorsten Wengler
  \medskip}

\address{ Physikalisches Institut, Universit\"at Heidelberg, 
  Philosophenweg 12, \\69120 Heidelberg, Germany, E-mail: 
  Thorsten.Wengler@desy.de \\ H1 Collaboration} 

\maketitle
\abstracts{High transverse momentum {\piz}-mesons and charged
  particles are measured in deep inelastic e-p scattering events at
  low Bjorken-$x$ taken with the H1 detector at HERA. The production
  of high {\pt} particles is strongly correlated to the emission of
  hard partons in QCD and is therefore sensitive to the dynamics of
  the strong interaction. For the first time the measurement of single
  particles has been extended to the region of small angles w.r.t. the
  proton remnant (forward region). This region is expected to be
  particularly sensitive to QCD evolution effects in final
  states. Results are presented as a function of Bjorken-$x$ and
  {\xall}, the fraction of the incident proton's energy carried by the
  particle, and are compared to different QCD models.} 

\section{Introduction}

The positron-proton collider HERA has significantly increased the
available kinematical region for the study of Deep Inelastic
Scattering (DIS) towards large four-momentum transfer 
\Qsq ~(up to \Qsq ~$ \approx 10^4$ \GeVsq) and  small Bjorken-$x$ 
($x \approx 10^{-4}$). 
The hadronic final state in such events has been suggested to
be a very sensitive environment to disentangle different QCD
processes and to test the validity of various approximations. In
particular the region towards the proton remnant (forward region) is
considered to be among the most promising areas to gain new insights
into dynamical features of QCD at low Bjorken-$x$~\cite{padyn} since
it provides a large phase space for parton emissions. Several
prescriptions for the QCD evolution of the dynamics of such emissions
have been proposed. 
The DGLAP (Dokshitzer-Gribov-Lipatov-Altarelli-Parisi)~\cite{dglap}
evolution, where the relevant evolution parameter is log$(Q^2/Q_0^2)$,  
has been successfully tested over wide ranges in \Qsq, 
and provides, for instance, a good description of the structure
function scaling violations. The BFKL
(Balitsky-Fadin-Kuraev-Lipatov)~\cite{bfkl} evolution equation, where
the relevant evolution parameter is log$(1/x)$, is expected to become 
applicable at small enough $x$ where the log$(1/x)$ terms
dominate the evolution.
The CCFM (Ciafaloni-Catani-Fiorani-Marchesini)~\cite{ccfm} evolution,
where the parton emissions obey angular ordering, is expected to be
valid for both low and high Bjorken-$x$, thereby forming a bridge
between the BFKL and DGLAP approaches. Hadronic
final state calculations for all three prescriptions are, for the
first time, becoming available~\cite{newmc}.

In this analysis data from the H1 experiment are used to study 
high energetic {\piz}-meson
and charged particle production in the forward region at small
Bjorken-$x$. 
High {\pt} particles are strongly correlated to the production of hard
partons. They can therefore be used to penetrate screening
hadronization effects. Previous measurements~\cite{h1paspec}
exploiting this feature have been limited by experimental acceptance
mostly to the central region of the detector. The study of forward
going jets has also been suggested~\cite{padyn} to study parton
dynamics at low Bjorken-$x$. 
The integrating nature of jet algorithms and the spatial extension of
jets thus defined however make such observables liable to
contributions from the proton remnant.
The measurement of single particles provides complementary observables
independent of jet algorithms while taking full advantage of the
detector acceptance. Another interesting feature of single particle
observables is that they can be directly compared to analytical
calculations by employing fragmentation functions.
In particular the {\piz}-mesons can be measured to
high $\xpi = \epi/\epro$ (see below). For these values of 
{\xpi} the parton density functions in the proton are well
known from the global QCD analyses and small Bjorken-$x$ dynamics
can be exposed free from the ambiguities associated with the choice
of the non-perturbative parton input~\cite{fwdpi0}.

\section{Data Selection}

Experimental data for this analysis were collected by the H1 
experiment during the 1994
running period, in which HERA collided 27.5 GeV positrons on 820 GeV
protons. Integrated luminosities of the data samples used for the
 {\piz}-meson and charged particle measurements are 1.9 pb$^{-1}$ 
and 1.0 pb$^{-1}$ respectively. DIS events are selected 
in the region of Bjorken-$x$ 
~$2\cdot 10^{-4} < x < 2.36\cdot 10^{-3}$ ~via the
scattered positron satisfying $E_{e} > 12 $ GeV, 
$156^{\circ} < \Theta_{e} < 173^{\circ}$ and $y > 0.1$.
All results quoted here are corrected to this range.

\section{Forward $\pi^\circ$-mesons}

\begin{figure}[t]
    \vspace*{9.9cm}
    \includegraphics{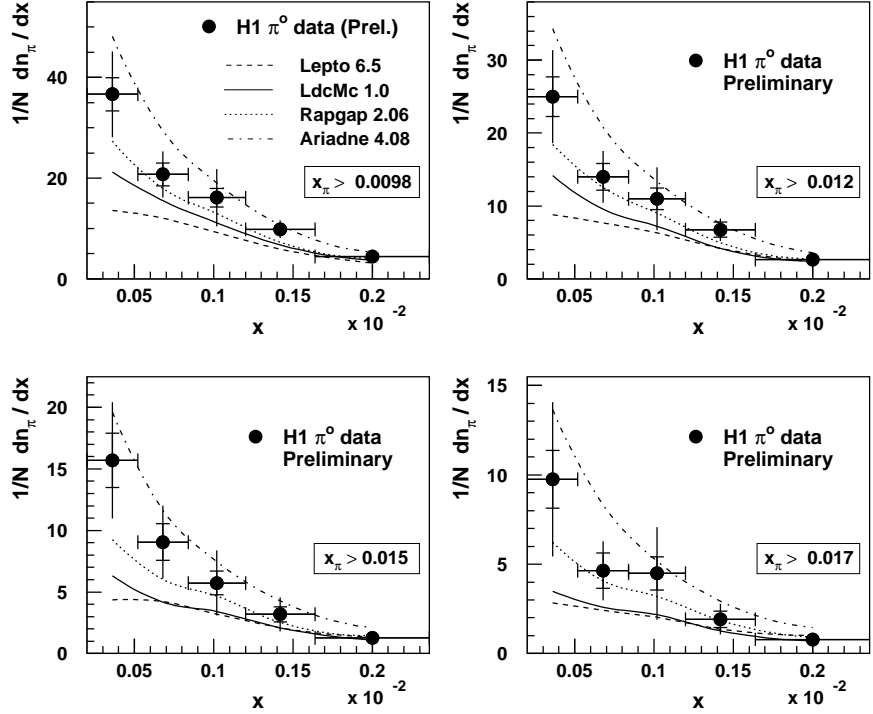}
\caption{The forward {\piz} spectra in Bjorken-$x$ for 
  {\ptpi} $>$ 1 GeV and ~$5^{\circ} < \thpi < 25^{\circ}$
  are shown for four different values of the lower threshold 
  {\xpi} = {\epi}/{\epro}. Here $n_{\pi}$ is the number of 
  {\piz}-mesons and $N$ is the number of events that enter the
  distribution. For comparison, the predictions of four different QCD
  models are overlaid.}
\label{pi0spec}
\end{figure}

The {\piz}-mesons are measured in the dominant decay channel 
{\piz} $\rightarrow 2\gamma$. The {\piz} candidates are selected in
the region of ~$5^{\circ} < {\thpi} < 25^{\circ}$ using the 
finely segmented H1 Liquid Argon (LAr) calorimeter, where {\thpi} is
the polar angle of the produced {\piz} with respect to the incoming
proton beam direction.  
Candidates are required to have an energy of {\epi} $>$ 8 GeV with
a transverse component {\etpi} = {\epi} sin{\thpi} $>$ 1 GeV.
At the high {\piz} energies considered here, the two photons cannot be
separated, but appear as one object (cluster) in the
calorimetric response. Photon induced showers are selected by
measuring the shower shapes of candidate clusters.  
The selection criteria are based on the compact nature 
of electromagnetic showers as opposed to those of hadronic origin.
The very fine segments ($\sim$~44000 cells in total, with a cell size
of 3.5 x 3.5 cm$^2$ and four-fold longitudinal segmentation in the
forward region) of the H1 LAr calorimeter allow for a detailed 
study of the transverse and longitudinal spread and the energy
distribution of each cluster. The main challenge of the analysis
is the high particle density in the forward direction leading to
overlapping showers of electromagnetic and hadronic origin.
Candidates of this type are mostly rejected. 
The same high multiplicity of particles makes the use of track
information as a rejection criterion infeasible for the {\piz} selection.
Monte Carlo studies using a detailed simulation of the H1 detector
nevertheless show the detection efficiency for {\piz}-mesons in the
region specified before to be above $40 \%$ and predict that more than
$75 \%$ of the selected candidate showers stem from {\piz}-mesons.
The contribution of candidates not associated to the primary vertex
is below $10 \%$. All distributions shown are corrected bin-by-bin for
these detector effects and QED radiation.

Results are presented as spectra in Bjorken-$x$ of {\piz}-mesons with 
~$5^{\circ} < \thpi < 25^{\circ}$ and ~{\ptpi} $>$ 1 GeV. The data is
shown in Fig. \ref{pi0spec} for the four different lower thresholds of
\mbox{{\xpi} $> \left[~ 0.0098, ~0.012, ~0.015, ~0.017 ~ \right]$}
to illustrate the dependence on {\xpi} = {\epi}/{\epro}.
The full errors are the quadratic sum of the
statistical (inner error bars) and systematic errors.
All distributions are normalized to the number of DIS events $N$ that
fall into the kinematic range specified before.
The data clearly rise with decreasing Bjorken-$x$, giving evidence of 
much more hard partonic radiation than predicted by
DGLAP type QCD models as represented by LEPTO~\cite{lepto}.
A better description of the data can be obtained by adding
contributions from processes in which the photon is not point-like but
acts as an resolved object as implemented in
RAPGAP~\cite{rapgap}. The choices of scales~\cite{jung} are however
somewhat arbitrary here and introduce
large uncertainties. For the first time, the data can also be
compared to a MC model based on the CCFM approach (LDCMC~\cite{newmc}). 
Although it is closer to the data for lower cutoffs in {\xpi} it becomes
increasingly similar to the DGLAP based model towards higher {\piz} energies.
The Color Dipole Model as implemented in ARIADNE~\cite{ariadne} tends
to overshoot the data in most of the measured distributions.

\section{Forward Charged Particles}

Charged particles (cp) are selected using the forward (FT) and central (CT)  
tracking chambers. Basic quality criteria are used to select well measured
tracks originating from the primary interaction point. 
For tracks restricted to the kinematic range used in this analysis: 
{\pcp} $>$ 8 GeV, {\ptcp} $>$ 1 GeV, $5^{\circ} < \thcp < 25^{\circ}$, 
over $90\%$ are reconstructed in the FT. For this measurement, as for
the {\piz}-mesons, the high particle multiplicity in the forward
direction constitutes the main challenge. It gives rise to a large
flux of soft particles produced in secondary interactions in passive
material within the detector leading to a degradation of detection
efficiency and resolution.
Since the curvature of the tracks is used to measure their
momentum then the resolution also decreases with increasing momentum. The
measurement of charged particles in this analysis is therefore only
currently feasible for the lowest threshold of 
$\xcp = \ecp/\epro > 0.0098$ used in the {\piz}-meson analysis.
The efficiency of reconstructing a central track and associating it to the 
primary event vertex is over 95$\%$~\cite{pvm}. The equivalent 
forward track efficiency is over 40$\%$~\cite{FT} for tracks well
contained within the FT. For tracks produced at low values of {\thcp}
\mbox{({\thcp} $< 10^\circ$)}, a lower efficiency of 25$\%$ is obtained. 
The proportion of these selected charged particles produced in secondary
interactions in passive material is negligible for central tracks but
is estimated to be maximally 10$\%$ for the forward sample used here. 
The data have been corrected bin-by-bin for the influence of the
detector and QED radiation. 

\begin{figure}[t]
    \vspace*{5.6cm}
    \includegraphics{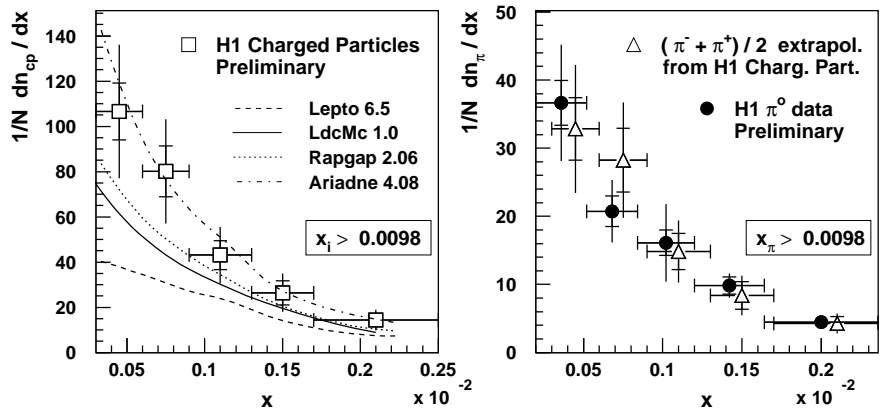}
\caption{The forward charged particle spectra in Bjorken-$x$ are shown
  for $\xcp = \ecp/\epro > 0.0098$ and $\ptcp > 1$ GeV
  in the polar angular range $5^{\circ} < \thcp < 25^{\circ}$.
  Here $n_{cp}$ is the number of 
  charged particles and $N$ is the number of events that enter the
  distribution. For comparison, the predictions of four different QCD
  models are overlaid.}
\label{chpart}
\end{figure}

The charged particle results are shown as spectra in Bjorken-$x$ in
\mbox{Fig. \ref{chpart}} (left) for the region defined above.
Both the data and the QCD models show the same Bjorken-$x$ behavior as
the {\piz}-meson spectra. To further check the consistency of both
measurements JETSET~\cite{string} has been used to estimate
the contribution of charged pions to the charged particle data.
The results divided by two are compared to the {\piz} measurement in
\mbox{Fig. \ref{chpart}} (right) and show good agreement as expected
from isospin symmetry.  

\section{Summary}

Two measurements have been presented which for the first time access
the experimentally difficult region close to the proton remnant for
the study of high {\pt} {\piz}-meson and charged particle production.
The data is presented as spectra in Bjorken-$x$.
The two analysis can be compared and are in good agreement.
In both cases the data clearly rise with decreasing Bjorken-$x$,
giving evidence of much more hard partonic radiation than predicted by
DGLAP type QCD models. The {\piz} measurement can access higher values
of {\xpi} and the discrepancy can be seen to increase there. Additional
contributions from processes with a resolved photon as implemented in
RAPGAP lead to a better description of both {\piz}-mesons and charged
particles but here large uncertainties are introduced since one is
forced to make a best guess of a reasonable scale~\cite{jung}.
For the first time, predictions for the hadronic final state from
a QCD model based on the CCFM approach (LDCMC) are available. 
Although they are closer to the data for lower cutoffs in {\xpi} they
become increasingly similar to the DGLAP based model towards higher
{\piz} energies. The Color Dipole Model as implemented in ARIADNE
tends to overshoot the data in most distributions.

\end{document}